\documentclass[11pt]{article}
\usepackage[margin=1.2in]{geometry}
\usepackage{enumitem}
\usepackage{amsmath}
\usepackage{listings}
\usepackage{hyperref}
\usepackage{graphicx}
\usepackage{hhline}
\usepackage{times}
\usepackage{bm}
\usepackage{afterpage}
\usepackage{comment}
\usepackage{amssymb}
\usepackage{indentfirst}
\usepackage[table]{xcolor}
\usepackage{xcolor}
\usepackage{verbatim}

\usepackage{siunitx}
\usepackage{tabularx,booktabs}
\usepackage{caption}
\usepackage[skip=3ex]{subcaption}

\usepackage{algorithm,algcompatible,amsmath}
\usepackage{setspace}
\usepackage[backend=bibtex, style=bwl-FU, maxcitenames=2]{biblatex}
\usepackage[font=small,textfont=it]{caption}
\algnewcommand\BLOCK{\item[\textbf{Block 1.}]}%
\algnewcommand\BLOCKK{\item[\textbf{Block 2.}]}%
\algnewcommand\BLOCKKK{\item[\textbf{Block 3.}]}%
\algnewcommand\BLOCKKKK{\item[\textbf{Block 4.}]}%

\algnewcommand\STEPA{\item[\textbf{Step (a)}]}%
\algnewcommand\STEPB{\item[\textbf{Step (b)}]}%
\algnewcommand\STEPC{\item[\textbf{Step (c)}]}%

\title{A Review of Bayesian Methods for Infinite Factorisations}

\date{\today}

\author{Margarita Grushanina\thanks{Department of Economics, Vienna University of Economics and Business, Welthandelsplatz 1, 1020 Vienna, Austria}}

\newcommand{\diag}{\ensuremath{\mathrm{diag}}}

\begin{document}
\maketitle

\begin{abstract}
\noindent

Defining the number of latent factors has been one of the most challenging problems in factor analysis. Infinite factor models offer a solution to this problem by applying increasing shrinkage on the columns of factor loading matrices, thus penalising increasing factor dimensionality. The adaptive MCMC algorithms used for inference in such models allow to defer the dimension of the latent factor space automatically based on the data. This paper presents an overview of Bayesian models for infinite factorisations with some discussion on the properties of such models as well as their comparative advantages and drawbacks.

\begin{keywords}
Factor analysis, adaptive Gibbs sampling, spike-and-slab prior, Indian buffet process, multiplicative gamma process, increasing shrinkage
\end{keywords}

\end{abstract}

\section{Introduction}

Latent factor models represent a popular tool for data analysis in many areas of science, including psychology, marketing, economics, finance, genetic research, pharmacology and medicine. Their history dates back to \cite{Spear1904}, who first suggested common factor analysis as a single factor model in the context of psychology. \cite{Thu1931} and \cite{Thu1934} extended it to multiple common factors and introduced some important factor analysis concepts, such as communality, uniqueness, and rotation. \cite{Anderson1956} in their seminal paper established important theoretical foundations of latent factor analysis. Since then there has been a vast and constantly growing pool of literature covering various theoretical and practical aspects of factor analysis. Some selective reviews include, for example, \cite{DynRev2013} and \cite{SW2016} for dynamic factor models, \cite{BW2016} for large factor models, and \cite{Fan2021} for factor models in application to econometric learning.

Recent years have also seen a considerable research in the area of Bayesian latent factor models. Some of the many important contributions in this area include \cite{GZh1996}, \cite{AW2000}, \cite{West2003}, \cite{Lopes2004}, \cite{SFSLopes2010}, \cite{Conti2014}, \cite{Rockova2016}, \cite{KaufSchu2019} and \cite{SFSDHHL2022}.

One of the most challenging tasks in factor analysis concerns the inference of the true number of latent factors in the model. The most common approach in the literature has long been to use various criteria to choose a model with the correct number of factors. Thus, \cite{Bai2002} use information criteria to compare models with different factors' cardinalities. \cite{Kapetanios2010} performs model comparison using test statistics, while \cite{Polasek1997} and \cite{Lopes2004} rely on marginal likelihood estimation to determine the true number of factors in the model. \cite{Carvalho2008} perform the evolutionary stochastic model search which iteratively increases the model by an additional factor until reaching some pre-specified limit or until the process stops including additional factors. As a different approach, \cite{Lopes2004} customise a reversible jump MCMC (RJMCMC) algorithm introduced in \cite{Green1995} for moving between models with different numbers of factors, while \cite{SFSLopes2018} suggest a one-sweep algorithm to estimate the true number of factors from an overfitting factor model.

However, such methods are often computationally demanding, especially when the dimensionality of the analysed data set is high. Recently, another approach has been developed which allows the factors' cardinality to be derived from data by letting the number of factors to potentially be infinite. The dimension reduction is then achieved by assigning a nonparametric prior to factor loadings which penalises the increase of the number of columns in the factor loading matrix via increasing shrinkage of the factor loadings on each additional factor to zero. Thus, in their pioneering work, \cite{BD2011} introduced the multiplicative gamma process (MGP) prior on the precision of factor loadings, which is defined as a cumulative product of gamma distributions. \cite{Knowles2011} and \cite{Rockova2016} employed the Indian Buffet Process (IBP) to enforce sparsity on factor loadings and at the same time penalise the increasing dimensionality of latent factors. \cite{LDD2020} introduced the cumulative shrinkage process (CUSP) prior which applies cumulative shrinkage on the increasing number of columns of the factor loading matrix via a sequence of spike-and-slab distributions. Model inference is usually performed via Gibbs sampler steps, however, the models' changing dimensions at different iterations of the sampler require the usage of adaptive algorithms, which have some specific properties that need to be taken into account.

This paper provides a review of the methods for infinite factorisations, with a focus on their properties, comparative advantages and drawbacks. The paper proceeds as follows: Section \ref{BFA} briefly reviews the formulation of a Bayesian factor model and a shrinkage prior on factor loadings. Sections \ref{MGP} - \ref{IBP} provide an insight into the three above mentioned priors for infinite factorisations, namely, MGP, CUSP and IBP priors, and outline their main advantages and drawbacks. Section \ref{gen:inf:fac} reviews the concept of generalized infinite factorization models. Section \ref{Conclusion} concludes with a discussion.

\section{Bayesian infinite factor model} \label{BFA}

	\subsection{Bayesian latent factor model}   \label{linBFA}
	
	In the traditional Bayesian factor analysis data on $p$ related variables are assumed to arise from a multivariate normal distribution $\bm{y}_t \sim N_p (\bm{0},\bm{\Omega})$, where $\bm{y}_t$ is the $t$-th of the $T$ observations and $\bm{\Omega}$ is the unknown covariance matrix of the data. A factor model represents each observation $\bm{y}_t$ as a linear combination of $K$ common factors $\bm{f}_t = (f_{1t}, \ldots, f_{Kt})^T$:
\begin{align} \label{eq:factors}
\bm{y}_t = \bm{\Lambda f}_t + \bm{\epsilon}_t ,
\end{align}
where $\bm{\Lambda}$ is an unknown $p \times K$ factor loading matrix with factor loadings $\lambda_{ih}$ ($i =1, \ldots, p$, $h=1, \ldots, K$) and it is typically assumed that $K \ll p$.

Often, the latent factors are assumed to be orthogonal and follow a normal distribution $\bm{f}_t \sim N_p(\bm{0}, \bm{I}_p)$. Furthermore, it is assumed that the factors $\bm{f}_t$ and $\bm{f}_s$ are pairwise independent for $t \neq s$. The idiosyncratic errors $\epsilon_t$ are also assumed normal and pairwise independent:
\begin{align*} %\label{lab:eq2}
        \epsilon_t \sim N_{p}(\bm{0}, \bm{\Sigma}),  \quad \quad  \bm{\Sigma}  = \diag(\sigma_1^2, \ldots, \sigma_{p}^2) .
\end{align*}

These assumptions allow to represent the covariance matrix of the data in the following way: 
\begin{align} \label{eq:cov}
\bm{\Omega}  = \bm{\Lambda \Lambda}^T + \bm{\Sigma} .
\end{align}

There are many different ways to choose a prior for the elements of the factor loading matrix $\bm{\Lambda}$. A typical choice involves a version of a normal prior $\lambda_{ih} \sim N(d_{ih}^0, D_{ih}^0)$ for the reason of conjugacy. The hyperparameter $d_{ih}^0$ is often chosen to be equal to zero. This has an additional advantage that with a suitably chosen hyperprior for $D_{ih}^0$ such setting can result in a sparse $\bm{\Lambda}$ with many zero elements, which is justified for many applications of factor models. To ensure identifiability, it is often assumed that $\bm{\Lambda}$ has a full rank lower triangular structure, which imposes a choice of a truncated normal prior for the diagonal elements of $\bm{\Lambda}$ to ensure positivity and a normal prior for the lower diagonal elements (see, e.g. \cite{GZh1996}, \cite{Lopes2004}, \cite{GD2009}, amongst others).

The idiosyncratic variances $\sigma_i^2$ are usually assigned an inverse Gamma prior $\sigma_i^2 \sim \mathcal{G}^{-1}(c_{0i}, C_{0i})$ mainly for the reasons  of its conditional conjugacy.

	\subsection{Standard Gibbs sampler} \label{Gibbs:standard}

Inference is usually performed via a Gibbs sampler, sequentially sampling factor loadings, idiosyncratic variances and factors from their respective conditional distributions. These steps are rather generic for a wide range of factor models and choices of parameters. Assuming that the data is explained by $K$ latent factors and that in the normal prior for the elements of the factor loading matrix $d_{ih}^0 = 0$, the Gibbs sampler steps for updating $\bm{\Lambda}$, $\bm{\Sigma}$ and $\bm{F} = \{ \bm{f}_t:  t = 1, \ldots, T \}$ will look as follows:

\vspace*{2mm}

\noindent
\textit{Step 1.} Sample $\bm{\lambda}_i$ for $i$ in $(1, \ldots, p)$ from
\begin{align*}
 \bm{\lambda}^T_i|- \sim N_{K} \left((\bm{\Psi}_i^{-1} + \sigma_i^{-2}\bm{FF^{T}})^{-1}\bm{F}\sigma_i^{-2}\bm{y}_i^{T}, (\bm{\Psi}_i^{-1} + \sigma_i^{-2}\bm{FF}^{T})^{-1} \right)
\end{align*}
where $\bm{\Psi}_i = \diag(D_{i1}^0, \ldots, D_{iK}^0)$ and $\bm{\lambda}_i$ is the $i$th row of the factor loading matrix $\bm{\Lambda}$.

\vspace*{2mm}

\noindent
\textit{Step 2.} Sample $\sigma_i^{-2}$ for $i$ in $(1, \ldots, p)$ from
\begin{align*}
 \sigma_i^{-2}|- \sim \mathcal{G} \left(c_{0i}+\frac{T}{2},C_{0i}+\frac{1}{2}\sum_{t=1}^{T}(y_{it}-\bm\lambda^T_i\bm{f}_t)^2\right) .
\end{align*}

\vspace*{2mm}

\noindent
\textit{Step 3.} Sample $\bm{f}_t$ for $t$ in $(1, \ldots, T)$ from
\begin{align*}
 \bm{f}_t|- \sim N_{K} \left((\bm{I}_{K} + \bm{\Lambda}^T\bm{\Sigma}^{-1}\bm{\Lambda})^{-1}\bm{\Lambda}^T\bm{\Sigma}^{-1}\bm{y}_t, (\bm{I}_{K} + \bm{\Lambda}^T\bm{\Sigma}^{-1}\bm{\Lambda})^{-1} \right)
\end{align*}
where $\bm{\Sigma} = \diag(\sigma_{1}^{2}, \ldots, \sigma_{p}^{2})$.

Additional steps can be added to update hyperparameters if hyperpriors are assigned to any of the parameters of the prior distributions for $\lambda_{ih}$ and $\sigma_i^2$.

	\subsection{Infinite factorisations and increasing shrinkage of the prior for $\mathbf{\Lambda}$}

	In the above described Gibbs sampler steps, we take the number of latent factors $K$ as known. In reality, this is rarely the case and determining the plausible number of latent factors can be a difficult and time consuming problem, especially in high-dimensional data sets. In the last decade, there has been a rise in the literature using a different approach towards determining the number of latent factors. This approach assumes that a factor model can in theory include infinitely many factors, i.e. the factor loading matrix $\bm{\Lambda}$ can be comprised of infinitely many columns. This means that $\bm{\Lambda}$ is seen as a parameter-expanded factor loading matrix with redundant parameters.

More formally, if $\bm{\Theta}_\Lambda$ denotes the collection of all matrices $\bm{\Lambda}$ with $p$ rows and infinitely many columns, then the product $\bm{\Lambda} \bm{\Lambda}^T$ is a $p \times p$ matrix with all entries finite if and only if the following condition holds\footnote{This follows from the Cauchy-Schwartz inequality, see the proof in \cite{BD2011}.}:
\begin{align*}
\bm{\Theta}_\Lambda = \Bigl\{ \bm{\Lambda} = (\lambda_{ih}), \: i=1, \ldots, p, \: h=1, \ldots, \infty, \: \max_{1 \leq i \leq p} \sum_{h=1}^\infty \lambda_{ih}^2 < \infty \Bigr\}
\end{align*}
	
The prior on the elements of $\bm{\Lambda}$ is defined in such a way that it allows $\lambda_{ih}$s to decrease in magnitude if the column index $h$ grows, thus penalising the increasing factor dimensionality. This approach allows the number of factors to be derived automatically from data via an adaptive inference algorithm. In the next sections we discuss the most notable methods for infinite factorisations in detail.

\section{Multiplicative gamma process prior} \label{MGP}

  \subsection{The prior specification}

In their seminal paper, \cite{BD2011} proposed one way to choose a prior on the elements of a factor loading matrix so that to penalise the effect of additional columns: $\lambda_{ih}$s are given a normal prior centred at zero, while the prior precisions of $\lambda_{ih}$s for each $h$ are defined as a cumulative product of gamma priors.

The MGP prior can be formalised as follows:
\begin{eqnarray} \label{eq:MGP}
   &     \lambda_{ih} \rvert \phi_{ih},\tau_{h}   \sim N(0, \phi_{ih}^{-1} \tau_{h}^{-1}),  \quad \quad  \phi_{ih} \sim \mathcal{G}(\nu_1/2, \nu_2/2),   \quad \quad  \tau_h = \displaystyle\prod_{l=1}^{h} \delta_l, & \\
&\delta_{1} \sim \mathcal{G}(a_1, b_1),   \qquad  \delta_{l} \sim \mathcal{G}(a_2, b_2),   \quad  l\geq2 , & \nonumber
 \end{eqnarray}
 where
$\delta_l$ $(l = 1, \ldots, \infty)$ are independent, $\tau_h$ is a global shrinkage parameter for the $h$-th column, $\phi_{ih}$ are local shrinkage parameters for the elements of the $h$-th column. The condition $a_2 > 1$ is imposed on the shape parameter of the prior for $\delta_l$ to insure that  $\tau_h$s are stochastically increasing with increasing $h$. In \cite{BD2011}, $b_1 = b_2 = 1$ are set at $1$, while  $a_1$ and $a_2$ are assigned the hyperprior $\mathcal{G}(2,1)$ and sampled in a Metropolis-within-Gibbs step.

	\subsection{Inference and adaptive Gibbs sampler} \label{GibbsMGP}
	
The inference is done via a Gibbs sampler with a few additional steps to the standard ones described in Section \ref{Gibbs:standard}. A distinctive feature of the sampler suggested in \cite{BD2011} is that it truncates the factor loading matrix  $\bm{\Lambda}$ to have $k^*$ columns, where $k^*$ is the number of factors supported by the data at each given iteration of the sampler. The truncation procedure deserves some closer attention.

Although theoretically the number of factors is allowed to be infinitely large, in reality one chooses a suitable level of truncation $k^*$, designed to be large enough not to miss any important factors, but also not too conservative to induce unnecessary computational effort. The sampler is initiated with a conservative guess $K_0$, which is chosen to be substantially larger than the supposed actual number of factors. At each iteration of the sampler, the posterior samples of the factor loading matrix $\bm{\Lambda}$ contain information about the effective number of factors supported by the data in the following way. Let $m^{(g)}$ be the number of columns of $\bm{\Lambda}$ at iteration $g$ which have all their elements so small that they fall within some pre-specified neighbourhood of zero. Then these columns are considered redundant and $k^{*(g)} = k^{*(g-1)} - m^{(g)}$ is defined to be the effective number of factors at iteration $g$. To keep balance between dimensionality reduction and exploring the whole space of possible factors, $k^*$ is adapted with probability $p(g) = \exp(\alpha_0 + \alpha_1 g)$, with the parameters chosen so that the adaptation occurs more often at the beginning of the chain and decreases in frequency exponentially fast (the adaptation is designed to satisfy the diminishing adaptation condition in Theorem 5 of \cite{Roberts2007}, which is necessary for convergence). When the adaptation occurs, the redundant factors are discarded and the corresponding columns are deleted from the loading matrix (together with all other corresponding parameters). If none of the columns appear redundant at iteration $g$, a factor is added, with all its parameters sampled from the corresponding prior distributions. Adaptation is made to occur after a suitable burn-in period in order to ensure that the true posterior distribution is being sampled from before truncating the loading matrices.

In the adaptive Gibbs sampler with the MGP prior on the factor loadings, the first three steps will be essentially the same as in Section \ref{Gibbs:standard}, with two alterations: the number of factors $K$ will be replaced by $k^*$ and in Step $1$ $D^0_{i1}, \ldots, D^0_{iK}$ will consequently be replaced by $\phi_{i1}^{-1} \tau_1^{-1}, \ldots, \phi_{ik^*}^{-1} \tau_{k^*}^{-1}$. The additional steps will have the following form:

\vspace*{2mm}

\noindent
\textit{Step 4.} Sample $\phi_{ih}$ for $i$ in $(1, \ldots, p)$ and $h$ in $(1, \ldots, k^*)$ from
\begin{align*}
 \phi_{ih}|- \sim \mathcal{G} \left(\frac{\nu_1 + 1}{2},\frac{\nu_2 + \tau_h \lambda_{ih}^2}{2}\right) \ .
\end{align*}

\vspace*{2mm}

\noindent
\textit{Step 5.} Sample $\delta_1$ from
\begin{align*}
 \delta_{1}|- \sim \mathcal{G} \left(\frac{2a_1 + pk^*}{2},1 + \frac{1}{2}\displaystyle\sum_{l=1}^{k^*}\tau_l^{(1)}\displaystyle\sum_{i=1}^{p}\phi_{il} \lambda_{il}^2 \right) \ .
\end{align*}
Sample $\delta_{h}$ for $h \geq 2$ from
\begin{align*}
 \delta_{h}|- \sim \mathcal{G} \left(\frac{2a_2 + p(k^* - h + 1)}{2},1 + \frac{1}{2}\displaystyle\sum_{l=h}^{k^*}\tau_l^{(h)}\displaystyle\sum_{i=1}^{p}\phi_{il} \lambda_{il}^2 \right)
\end{align*}
where $\tau_l^{(h)}= \prod_{t=1, t \neq h}^{l} \delta_t$ for $h$ in $(1, \ldots, k^*)$.

\vspace*{2mm}

\noindent
\textit{Step 6.} Sample the posterior densities of $a_1|\delta_1$ and $a_2|
\delta_2, \ldots, \delta_{k^*}$ via a random walk Metropolis-Hastings step with $a_1^p \sim N(a_1, s_1^2)$ and $a_2^p \sim N(a_2, s_2^2)$ serving as proposal quantities and the acceptance probabilities being:
\begin{eqnarray*}
&& \rho_{a_1} = \frac{\Gamma(a_1)}{\Gamma(a_1^p)} \: \frac{a_1^p}{a_1} \: \delta_1^{a_1^p - a_1} \: e^{a_1 - a_1^p},\\
&& \rho_{a_2} = \left(\frac{\Gamma(a_2)}{\Gamma(a_2^p)}\right)^{-(k^* - 1)} \: \frac{a_2^p}{a_2} \: \left(\prod_{l=2}^{k^*}\delta_l\right)^{a_2^p - a_2} \: e^{a_2 - a_2^p}.
\end{eqnarray*}

\vspace*{2mm}

\noindent
\textit{Step 7.} At each iteration generate a random number $u_g$ from $\mathcal{U}(0,1)$. If $u_g \leq p(g)$, check if any columns of the factor loading matrix $\bm{\Lambda}$ are within the pre-specified neighbourhood of $0$, and if this is so, discard the redundant columns and all its corresponding parameters. In the case when the number of such columns is zero, generate an additional factor by sampling its parameters from the prior distributions.

\subsection{Practical applications and properties} \label{simMGP}

The MGP prior has initially been developed for high-dimensional datasets with $p \gg T$ and a sparse covariance matrix structure, such as genes expression data. However, it acquired a wide-spread popularity and has been proved useful in various applications, see e.g. \cite{Montagna2012} and \cite{Rai2014}, amongst others. An application of particular interest is the infinite mixture of infinite factor analysers (IMIFA) model introduced in \cite{Murphy2019}, where the MGP prior was used in the context of a mixture of factor analysers to allow automatic inference on the number of latent factors within each cluster.

However, the MGP model has also some important limitations. Some of these limitations are investigated in \cite{Durante2017}, who addressed the dependence of the shrinkage induced by the MGP prior on the value of the hyperparameters $a_1 > 0$ and $a_2 > 0$. \cite{BD2011} state that the $\tau_h$s in (\ref{eq:MGP}) are stochastically increasing with increasing $h$ under the restriction $a_2 > 1$, which means that the induced prior on $1/\tau_h$ increasingly shrinks the underlying quantity towards zero as the column index $h$ increases, provided that $a_2 > 1$. \cite{Durante2017} argues that this is not sufficient to guarantee the increasing shrinkage property in a general case. Instead, further conditions are required, such as 
\begin{align} \label{cond:Durante}
a_2 > b_2+1, \qquad a_2 > a_1 
\end{align}
for the increasing penalization of a high number of factors to hold (in expectation), providing that  $a_1 > 0$ and $a_2 > 0$ and the values of $a_1$ are not excessively high. In his simulation study of the performance of the MGP prior for various values of the hyperparameters $a_1$ and $a_2$, \cite{Durante2017} investigates the behaviour of the model with $T=100$, $p=10$, and two different values for the true number of factors, namely $K=2$ and $K=6$. The results show an improved posterior concentration when the parameters $a_1$ and $a_2$ satisfy condition (\ref{cond:Durante}), specially for the case $K=2$. As the true rank of the model increases, there is evidence that the shrinkage induced by the MGP prior might be too strong.

Another critique of the MGP prior appeared in \cite{LDD2020}, who pointed out that the hyperparameters $a_1$ and $a_2$ both control the rate of shrinkage and the prior for the loadings on active factors. This creates a trade-off between the need to maintain considerably diffuse priors for active components and the endeavour to shrink the redundant ones. In their simulation study, \cite{LDD2020} found that the MGP prior significantly overestimates the number of active factors on a medium sized data set with $p < T$.

In an attempt to evaluate the performance of the MGP prior when the hyperparameters $a_1$ and $a_2$ are derived from data, we simulated a dataset in a similar way as in \cite{BD2011}. More specifically, a synthetic data set was simulated with $T=100$ and idiosyncratic variances sampled from $\mathcal{G}^{-1}(1,0.25)$. The number of non-zero elements in each column of $\bm{\Lambda}$ were chosen between $2k$ and $k+1$, with zeros allocated randomly and non-zero elements sampled independently from $N(0,9)$. We generated $\bm{y}_t$ from $N_p(0, \bm{\Omega})$, where $\bm{\Omega}  = \bm{\Lambda \Lambda'} + \bm{\Sigma}$. Further, we chose six $(p,K)$ combinations to test various dimensions of $\bm{\Lambda}$, namely $(6, 2)$, $(10, 3)$, $(30, 5)$, $(50, 8)$, $(100, 15$) and $(150, 25)$ with a conservative initial upper bound of $k_0 = \min (p, 5 \log(p))$, and $k_0 = 10\log(p)$ for the latter case with $p > T$. For each pair we considered 10 simulation replicates. The simulation was run for $30 000$ iterations with a burn-in of $10 000$. 

We used the following hyperparameter values: $\nu_1$ and $\nu_2$ both equal to $3$, the rate parameters $b_1$ and $b_2$ in the Gamma priors for $\delta_1$ and $\delta_l$ are set at $1$. For the case when $p < T$, $\alpha_0$ and $\alpha_1$ in the adaptation probability expression were set as $-0.5$ and $-3 \times (10)^{-4}$, and as $-1$ and $-5 \times (10)^{-4}$ for the case when $p \geq T$. The threshold for monitoring the columns to discard as $0.01$\footnote{Setting the threshold for monitoring the redundant columns at a smaller value than $0.01$ in the case when $p \geq T$ led to a an improvement of the results. However, tuning the threshold parameters remains highly heuristic and can be tricky while working with real data sets when the true number of factors is not known.} with the proportion of elements required to be below the threshold at $80$ \% of $p$.

The simulation results in Table \ref{tab:sim1} show that the model tends to overestimate the number of active factors in the case when $p \leq T$. In the last case, when the number of variables $p$ exceeds the number of observations $T$, the number of active factors is severely underestimated compared to the true one. The last two columns in Table~\ref{tab:sim1} show the posterior mean of $a_1$ and $a_2$. The first efficient shrinkage condition of \cite{Durante2017}, $a_2 > b_2+1$, holds for all $(p, k)$ combinations considered. For the first three combinations of $p$ and $k$, the column shrinkage parameters $a_1$ and $a_2$, estimated from the data, are in accordance with the second efficient shrinkage condition of \cite{Durante2017}, namely $a_2 > a_1$. However, with higher $p$, the condition $a_2 > a_1$ seems to cease holding when $p$ gets closer to $50$. This result is of some interest especially in view of the simulation study in \cite{Durante2017}, which suggests that the shrinkage induced by the MGP prior (and satisfying the condition $a_2 > a_1$) might prove too strong when the dimension of the data set increases.

\begin{table}[t!]
  \begin{center}
    \begin{tabular}{c|c|c|c|c}
    \hline
      \text{$(p,K)$} & \text{mode $k^*$} & \text{IQR} & \text{$\hat{a}_1$} & \text{$\hat{a}_2$}\\
      \hline
      $(6, 2)$ & 6.00 & 1.00 & 1.41 & 5.89\\
      $(10, 3)$ & 5.75 & 1.30 & 1.31 & 5.12\\
      $(30, 5)$ & 8.34 & 1.30 & 2.61 & 3.27\\
      $(50, 8)$ & 12.30 & 1.60 & 2.62 & 2.49\\
      $(100, 15)$ & 19.80 & 1.70 & 2.68 & 2.10\\
      $(150, 25)$ & 5.00 & 0.00 & 4.32 & 4.96\\
      \hline
    \end{tabular}
  \end{center}
  \vspace{-5mm}
  \caption{Performance of the adaptive Gibbs sampler based on the MGP prior for various combinations of $p$ and $K$. The modal estimates of $k^*$ and the interquartile range (IQR) are reported. $\hat{a}_1$ and $\hat{a}_2$ are the estimates of the values of $a_1$ and $a_2$ in (\ref{eq:MGP}) inferred via the Metropolis-Hastings step.}
   \label{tab:sim1}
\end{table}

Assigning a hyperprior to influential parameters, like we did in the case of $a_1$ and $a_2$, is a good way to reduce uncertainty and subjectivity of the model. However, the adaptation mechanism of such a sampler involves several hyperparameters, which may need to be adjusted depending on the nature and dimensionality of data. For example, we used an additional parameter indicating the proportion of the factor loadings in the column of $\bm{\Lambda}$ which needs to be within the chosen neighbourhood of zero to be considered redundant. This was first introduced in \cite{Murphy2019}, who found the choice of these truncation parameters to be a delicate issue which strongly depends on the type of the data. The threshold defining the neighbourhood of $0$, which is used to decide which factor loadings should be discarded, is another such example. Moreover, the parameters of the adaptation probability, $\alpha_0$ and $\alpha_1$, also need some tuning. In our simulation study, the speed of the adaptation differed for the settings with $p < T$ and $p > T$, when using the same values for $\alpha_0$ and $\alpha_1$. 

The importance and difficulty of choosing a suitable truncation criteria in the adaptive infinite factor algorithms was addressed in \cite{Schiavon2020}. The authors argue that the choice of truncation criteria, such as the predefined neighbourhood of zero, plays a vital role for the performance of the model. The optimal value of the criterion depends of the scale of data, while the number of active factors can be severely underestimated if the value of the truncation criterion is too large, and severely overestimated if it is too small. This is especially true for high-dimensional data, as with $p$ getting larger, the probability of having all values of $|\lambda_{ih}|$ smaller than the predefined threshold goes to zero exponentially. In the absence of any guidance towards choosing an optimal value of such a threshold, this remains a highly subjective and random procedure. \cite{Schiavon2020} suggest another way to define a criterion for truncating the redundant factors, which is robust to the scale of the data and has a well-defined upper bound. The main idea is to truncate $\bm{\Lambda}$ in such a way that the truncated model is able to explain at least a fraction $Q \in (0, 1)$ of the total variability of the data, where the variability of $\bm{y}$ is measured by the trace of the covariance matrix $\bm{\Omega}$:
\begin{align*}
\frac{tr(\bm{\Lambda}_{k^*}\bm{\Lambda}_{k^*}^T) + tr(\bm{\Sigma})}{tr(\bm{\Omega})} \geq Q,
\end{align*}
where $\bm{\Lambda}_{k^*}$ denotes the factor loading matrix obtained by discarding the columns of $\bm{\Lambda}$ starting from $k^* + 1$. The authors conduct a simulation study which shows that using the suggested method to select the relevant active factors drastically improves the performance of the MGP model.

\section{Cumulative shrinkage process prior} \label{CUSP}

	\subsection{The prior specification} \label{priorCUSP}

\cite{LDD2020} proposed another type of a nonparametric prior on the variances of the elements of $\bm{\Lambda}$, which largely corrects the drawbacks of the MGP prior. The CUSP prior on the factor loadings induces shrinkage via a sequence of spike-and slab distributions that assign growing mass to the spike as the model complexity grows. The CUSP prior formalises as follows:
\begin{align*}
\lambda_{ih}\,|\,\theta_h \sim N(0, \theta_h), \quad \text{where} \:\, i = 1,\ldots, p \:\, \text{and} \:\,  h = 1, \ldots, \infty
\end{align*}
\begin{align} \label{eq:theta}
\theta_h\,|\,\pi_h \sim (1-\pi_h)\mathcal{G}^{-1}(a_{\theta}, b_{\theta}) + \pi_h \delta_{\theta_{\infty}}, \quad \quad \pi_h = \sum_{l=1}^h w_l, \quad \quad w_l = v_l \prod_{m=1}^{l-1}(1-v_m)
\end{align}
where $\pi_h \in (0,1)$ and the $v_h$s are generated independently from $\mathcal{B}(1, \alpha)$, following the usual stick-breaking representation introduced in \cite{set:con}.
By integrating out $\theta_h$, each loading $\lambda_{ih}$ has the marginal prior\footnote{In the equation (\ref{eq:theta}) the inverse gamma distribution for the slab is chosen for the reasons of conjugacy. In principle, this expression provides a general prior, where a sufficiently diffuse continuous distribution needs to be chosen for the slab.}
\begin{align*}
\lambda_{ih} \sim (1-\pi_h)t_{2a_{\theta}}(0, b_{\theta}/a_{\theta}) + \pi_h N(0,\theta_{\infty})
\end{align*}
where $t_{2a_{\theta}}(0, b_{\theta}/a_{\theta})$ denotes the Student-\textit{t} distribution with $2a_{\theta}$ degrees of freedom, location 0 and scale $b_{\theta}/a_{\theta}$. To facilitate effective shrinkage of the redundant factors, $\theta_{\infty}$ should be set close to 0. The authors recommend a small value $\theta_{\infty} > 0$, following \cite{IR2005}, as it induces a continuous shrinkage prior on every factor loading, thus improving mixing and identification of inactive factors. The authors use the fixed value of $\theta_{\infty} = 0.05$, however, it can be replaced by some continuous distribution without affecting the key properties of the prior. This is shown in \cite{Kowal2021}, where a normal mixture of inverse-gamma priors is employed for the spike and slab distributions. The slab parameters $a_{\theta}$ and $b_{\theta}$ should be specified so as to induce a moderately diffuse prior on active loadings. 

	\subsection{Inference and adaptive Gibbs sampler}

The inference is done via Gibbs sampler steps. Similarly to the MGP model, the first three steps remain essentially the same as in Section \ref{Gibbs:standard}, with the difference that in Step $1$ $D^0_{i1}, \ldots, D^0_{iK}$ will be replaced by $\theta_{1} \ldots, \theta_{H}$, where $H$ is the truncation level. This truncation level is chosen differently than in \cite{BD2011} and the adaptation process is also different and designed in such a way that it depends less on heuristically chosen parameters.

While the probability of adaptation at iteration $g$ of the sampler is also set to satisfy the diminishing adaptation condition of \cite{Roberts2007}, there is no need to pre-specify an ad-hoc parameter describing some small neighbourhood of $0$.  The inactive  columns of $\bm{\Lambda}$ are identified as those which are assigned to the spike and are discarded at iteration $g$ with the probability $p(g) = e^{\alpha_0 + \alpha_1 g}$ together with all corresponding parameters. If at iteration $g$ all columns of the factor loading matrix are identified as active, i.e. assigned to the slab, an additional column of $\bm{\Lambda}$ is generated from the spike and all the corresponding parameters are sampled from their respective prior distributions. The initial number of columns $H$ at which the CUSP model is truncated is set equal to $p+1$, following the consideration that there can be at most $p$ active factors and by construction at least one column is assigned to the spike. The assignment of the columns of $\bm{\Lambda}$ to spike or slab at iteration $g$ is done using $H^{(g)}$ categorical variables $z_{h} \in \{1,2, \ldots, H^{(g)}\}$ with a discrete prior $Pr(z_{h} = h \, | \, w_{h}) = w_{h}$, where $H^{(g)}$ is the number of columns in $\bm{\Lambda}$ at iteration $g$.

The additional Gibbs sampler steps will look as follows:

\vspace*{2mm}

\noindent
\textit{Step 4.} Sample $\theta_h$ in a data augmentation step. Thus, (\ref{eq:theta}) can be obtained by marginalising out independent latent indicators $z_h$ with probabilities $p(z_h=l \,|\,w_l) = w_l$ for $l = 1, \ldots,H$, from the equation
\begin{align*}
\theta_h \,|\,z_h \sim \{1-\mathbf{1}(z_h \leq h)\}\mathcal{G}^{-1}(a_\theta, b_\theta) + \mathbf{1}(z_h \leq h)\delta_{\theta_\infty}.
\end{align*}
Sample $z_h$ for $h$ in $(1, \ldots, H)$ from a categorical distribution with probabilities as below
\begin{align*}
 p(z_h = l \,|\, -) \sim
 \begin{cases}
      w_l N_p (\bm{\lambda}_h;0,\theta_\infty \bm{I}_p), \quad \quad \quad \quad &l = 1,\ldots,h,\\
      w_l t_{2a_\theta} \left(\bm{\lambda}_h;0,(b_\theta/a_\theta) \bm{I}_p \right), \quad \quad &l = h+1,\ldots,H.
    \end{cases}
\end{align*}

\vspace*{2mm}

\noindent
\textit{Step 5.} Sample $v_l$ for $l$ in $(1, \ldots, H-1)$ from
\begin{align*}
 v_l\,|\,- \sim \mathcal{B} \left(1+\sum_{h=1}^{H}\mathbf{1}(z_h=l), \alpha + \sum_{h=1}^{H}\mathbf{1}(z_h>l)  \right) \ .
\end{align*}
Set $v_H=1$ and update $w_1,\ldots,w_H$ from $w_l=v_l\prod_{m=1}^{l-1}(1-v_m)$.

\vspace*{2mm}

\noindent
\textit{Step 6.} For $h$ in $(1, \ldots, H)$:

if $z_h \leq h$ set $\theta_h = \theta_\infty$, otherwise sample $\theta_h$ from $\mathcal{G}^{-1}\left(a_\theta + \frac{1}{2}p, b_\theta + \frac{1}{2}\sum_{j=1}^{p}\lambda_{ih}^{2}\right).$

\vspace*{2mm}

\noindent
\textit{Step 7.} After some burn-in period $\tilde{g}$ required for the stabilization of the chain, the truncation index $H^{(g)}$ and the number of active factors $H^{*(g)} = \sum_{h=1}^{H^{(g)}}\mathbf{1}(z_h^{(g)} > h)$ are adapted with probability $p(g) = exp(\alpha_0 + \alpha_1g)$\footnote{The coefficients $\alpha_0$ and $\alpha_1$ are chosen according to the criteria described in Section~\ref{GibbsMGP}} as follows:
\begin{itemize}
	\item if $H^{*(g)} < H^{(g-1)}-1$:
		\begin{itemize}
                \item[]  set $H^{(g)} = H^{*(g)} + 1$, drop inactive columns in $\bm{\Lambda}^{(g)}$ along with the associated parameters in $\bm{F}^{(g)}$, $\bm{\theta}^{(g)}$ and $\bm{w}^{(g)}$, and add the final component sampled from the spike to $\bm{\Lambda}^{(g)}$, together with the associated parameters in $\bm{F}^{(g)}$, $\bm{\theta}^{(g)}$ and $\bm{w}^{(g)}$ sampled from the corresponding priors
         \end{itemize}
	\item otherwise:
	\begin{itemize}
                \item[]  set $H^{(g)} = H^{(g-1)} + 1$ and add the final column sampled from the spike to $\bm{\Lambda}^{(g)}$, together with the associated parameters in $\bm{F}^{(g)}$, $\bm{\theta}^{(g)}$ and $\bm{w}^{(g)}$ sampled from the corresponding priors.
         \end{itemize}
\end{itemize}

	\subsection{Practical applications and properties}
	
	Since its introduction, the CUSP prior has been widely used in both theoretical studies and practical applications. The most notable of them include \cite{Kowal2021}, who employed the further generalised CUSP prior in the context of nonparametric functional bases; \cite{SFS2022}, who extended the CUSP prior to the class of generalized cumulative shrinkage priors with arbitrary stick-breaking representations which might be finite or infinite; \cite{Gu2023}, who applied the CUSP prior to infer the number of latent binary variables in the context of a Bayesian Pyramid (a multilayer discrete latent structure model for discrete data).
	
	In contrast to the MGP prior, the CUSP prior on factor loadings provides a clear separation in the parameters which control active factors and the shrinkage of the redundant terms. Thus, the shrinkage rate depends on $\alpha$ in a sense that smaller values of $\alpha$ enforce more rapid shrinkage and therefore smaller number of factors. The parameters $a_\theta, b_\theta$ of the inverse gamma prior for the slab control modelling of active factors (the inverse gamma prior can be replaced by another suitable continuous prior) and can be sampled from data in the spirit of the parameters $a_1$ and $a_2$ in the MGP model.
	
	To evaluate the comparative performance of the model with the CUSP prior on the data sets of various dimensionality, we simulated data sets in the same way as in Section \ref{simMGP}. The stick breaking parameter $\alpha$, which represents a prior expectation of the number of active factors in the dataset, was set to $5$ (as in \cite{LDD2020}). We also choose the same parameters of the slab distribution as in \cite{LDD2020}, namely $a_{\theta}=b_{\theta}=2$ and $\theta_{\infty}=0.05$. The parameters of the adaptation probability of the sampler $\alpha_0$ and $\alpha_1$ were set as $-1$ and $-5 \times (10)^{-4}$. The simulations were run for 15,000 iterations, with 5,000 discarded as burn-in, as convergence was achieved faster than in the case of the MGP prior. The simulation results are presented in Table \ref{tab:sim2} and show that the model was able to recover the correct number of factors in all considered cases. 
	
	\begin{table}[t!] 
  \begin{center}
    \begin{tabular}{c|c|c}
    \hline
      \text{$(p,K)$} & \text{mode $H^*$} & \text{IQR} \\
      \hline
      $(6, 2)$ & 2.00 & 0.00 \\
      $(10, 3)$ & 3.00 & 0.00 \\
      $(30, 5)$ & 5.00 & 0.00 \\
      $(50, 8)$ & 8.00 & 0.00 \\
      $(100, 15)$ & 15.00 & 0.00 \\
      \hline
    \end{tabular}
     \label{tab:sim:tableC1}
  \end{center}
   \vspace{-5mm}
   \caption{Performance of the adaptive Gibbs sampler
   based on the CUSP prior for various combinations of $p$ and $K$. The modal estimates of $H^*$ and the interquartile range (IQR) are reported.}
   \label{tab:sim2}
\end{table}
	
The CUSP model offers significant advantages compared to the MGP model by eliminating the very subjective and influential truncation threshold and decoupling the generation mechanism for active and redundant components. This results in much more robust estimations of the number of factors in data sets of various dimensions. In our experience, assigning some continuous distribution to $\delta_{\theta_\infty}$ and a hyperprior to $b_\theta$ can improve the performance, especially on a non-standardised data sets. The model provides poor uncertainty quantification with the sampler often being stuck in one (in most cases correct) value of $H^*$. This problem was addressed in \cite{Kowal2021} by extending the CUSP prior with a parameter expansion scheme which disperses the shrinkage applied to the factors.

\section{Indian buffet process prior} \label{IBP}

	\subsection{The prior specification}
	
Another, slightly different approach to modelling factor loading matrices involves Indian Buffet Process (\cite{Griffiths2006}), which defines a distribution over infinite binary matrices, to provide sparsity and a framework for inferring the number of latent factors in the data set. This approach was first suggested in \cite{Knowles2011} and is formally presented below.

First, a binary matrix $\bm{Z}$ is introduced whose elements indicate whether an observed variable $i$ has a contribution (non-zero loading) of factor $h$. Then the elements of $\bm{\Lambda}$ can be modelled in the following way:
\begin{align*}
\lambda_{ih} | z_{ih} \sim z_{ih} N(\lambda_{ih}; 0, \beta_h^{-1}) + (1 - z_{ih})\delta_0 (\lambda_{ih}),
\end{align*}	
where $\beta_h$ is a precision of the factor loadings in the $h$th column of $\bm{\Lambda}$ and $\delta_0$ is a delta function with a point-mass at $0$.

Thus, the factor loadings are modelled via a spike-and-slab distribution, however, differently from the CUSP prior, the separation into the spike and the slab is done not with a variance parameter but directly for the factor loadings  $\lambda_{ih}$ via an auxiliary binary indicator matrix. This allows a potentially infinite number of latent factors, i.e. $\bm{Z}$ has infinitely many columns of which only a finite number will have nonzero entries. If $\pi_h$ is a probability of a factor $h$ contributing to any of the $p$ variables, and $K$ is (for the moment the finite) number of latent factors, the IBP with the intensity parameter $\alpha_{IB}$ arises from the Beta-Bernoulli prior:
\begin{align*}
z_{ih} | \pi_h \sim Bernoulli (\pi_h), \qquad \pi_h | \alpha_{IB} \sim \mathcal{B} \left( \frac{\alpha_{IB}}{K}, 1 \right),
\end{align*}
by setting $K \rightarrow \infty$ and integrating out $\pi_h$.

	\subsection{Inference and adaptive Gibbs sampler}
	
The inference is done via a Gibbs sampler, of which the second and the third steps are the same as in Section \ref{Gibbs:standard}. The initial number of factors, which will define the dimensions of $\bm{\Lambda}$ and $\bm{Z}$ is chosen as some conservative number which clearly overfits any possible number of factors in the data set.
Step $1$ has the difference that not the $i$th row of the factor loadings matrix $\bm{\Lambda}$ but each element $\lambda_{ih}$ is sampled separately from the univariate normal distribution, if $z_{ih}=1$:

\vspace*{2mm}

\noindent
\textit{Step 1.} Sample $\lambda_{ih}$ for which $z_{ih}=1$ from
\begin{align*}
 \lambda_{ih}|- \sim N \left((\beta_h + \sigma_i^{-2} \bm{f}_h \bm{f}_h^{T})^{-1} \sigma_i^{-2} \bm{f}_h \bm{y}_i^{T}, (\beta_h + \sigma_i^{-2} \bm{f}_h \bm{f}_h^{T})^{-1} \right)
\end{align*}
where $\bm{f}_h$ is a vector of $t = 1, \ldots, T$ observations of factor $h$.

The precisions $\beta_h$ will be sampled in the following way:

\vspace*{2mm}

\noindent
\textit{Step 4.} Sampling $\beta_h$ providing it is given a gamma prior $\mathcal{G}(a_\beta, b_\beta)$
\begin{align*}
\beta_h \,|\,z_h, \lambda_{ih} \sim \mathcal{G}\left( a_\beta + \frac{\sum_{i=1}^p z_{ih}}{2}, b_\beta + \sum_{i,h}\lambda_{ih}^2 \right).
\end{align*}

The binary indicator $z_{ih}$ can be sampled using the fact that it is possible to calculate the posterior density of the ratio $\frac{p(z_{ih}=1|-)}{p(z_{ih}=0|-)}$ from the likelihood and prior probabilities and for every element there can be only two events, $z_{ih}=1$ or $z_{ih}=0$. This is done in the following way:

\vspace*{2mm}

\noindent
\textit{Step 5.} Sample binary indicator $z_{ih}$ using
\begin{align*}
 \frac{p(z_{ih}=1|-)}{p(z_{ih}=0|-)} \sim \frac{\sqrt{(\beta_h + \sigma_i^{-2} \bm{f}_h \bm{f}_h^T)^{-1}\beta_h} exp \left( \frac{1}{2}(\beta_h + \sigma_i^{-2} \bm{f}_h \bm{f}_h^T)^{-1} (\sigma_i^{-2} \bm{f}_h \bm{y}_i^T)^2 \right) m_{-i,h}}{T-1-m_{-i,h}}
\end{align*}
where $m_{-i,h}$ is the number of other variables for which factor $h$ is active, not counting variable $i$. 

\vspace*{2mm}

Although the binary matrix $\bm{Z}$ has infinitely many columns, only the nonzero ones contribute to the likelihood. However, one needs to take into account the zero columns too, as the number of factors can (and in many cases will) change at the subsequent iterations of the sampler. Let us denote $\kappa_i$ the number of columns of $\bm{Z}$ which contain $1$ only in row $i$, so it will contain information about the number of factors which are only active for the variable $i$\footnote{In terms of the Indian Buffet Process this means the number of new dishes customer $i$ tries.}. After the sampling step 5, $\kappa_i=0$ for any $i$ by design, so the new factors $\kappa_i$ are sampled in a separate MH step. Note that this is not a random walk MH step as the proposal densities are not symmetric.

\vspace*{2mm}

\noindent
\textit{Step 6.} Sample the number of new active factors $\kappa_i$ in a MH step with the following proposal density
\begin{align*}
\rho_{\kappa_i} = (2\pi)^{\frac{T\kappa_i}{2}} |\bm{M}|^{-\frac{T}{2}} exp \left(\frac{1}{2} \sum_t \bm{m}^T \bm{M} \bm{m} \right) \frac{Pois(\kappa_i; \alpha_{IB}/(p-1))}{Pois(\kappa_i; \alpha_{IB} \nu / (p-1))} ,
\end{align*}
where $\nu>0$ is a tuning parameter aimed at improving mixing, $\bm{M} = \sigma_i^{-2} \bm{\lambda}_{\kappa_i} \bm{\lambda}_{\kappa_i}^T + \bm{I}_{\kappa_i}$ with $\bm{\lambda}_{\kappa_i}$ denoting a $1 \times \kappa_i$ vector of the new elements of the factor loading matrix, and $\bm{m} = \bm{M}^{-1} \sigma_i^{-2} \bm{\lambda}_{\kappa_i} (y_{it} - \bm{\lambda}_i^T \bm{f}_t)$. Steps 5 and 6 are designed to be in one loop for $i=(1, \ldots, p)$, i.e. for each variable $i$, first, the indicator $z_{ih}$ is sampled for every $h$, and then the number of new factors for variable $i$ is sampled in the following step.

\vspace*{2mm}
	
\noindent
\textit{Step 7.} Assuming the gamma prior $\mathcal{G}(a_\alpha, b_\alpha)$, sample the IBP strength parameter $\alpha_{IB}$ from
\begin{align*}
\alpha_{IB} \,|\, \bm{Z} \sim \mathcal{G} \left(a_\alpha + K_+, b_\alpha + \sum_{j=1}^p \frac{1}{j} \right),
\end{align*}
where $K_+$ is the number of active factors for which $z_{ih}=1$ at least for one $i$.

	\subsection{Practical applications and properties}
	
The IBP prior coupled with a spike-and-slab distribution proved to be a useful approach to model sparse factor loadings and represents an alternative to implementing an increasing shrinkage on the columns of the factor loading matrix in terms of inferring the number of active factors. A somewhat related work was introduced earlier by \cite{RD2008} in the context of a nonparametric Bayesian factor regression model, where a sparse IBP prior was coupled with a hierarchical prior over factors. The authors did not assume independence of factors as in traditional factor analysis, and instead of a normal prior used a Kingman's coalescent prior which describes an exchangeable distribution over a countable set of factors.

The original model of \cite{Knowles2011} was further extended in \cite{Rockova2016}, where the authors couple the IBP prior on the binary indicators with a spike-and-slab LASSO (SSL) prior of the elements of $\bm{\Lambda}$. The SSL prior assigns to both the spike and the slab components a Laplace distribution designed so that the slab has a common scale parameter and the spike has a factor-specific scale parameter (different for each $h$). This prior tackles the problem of rotational invariance of $\bm{\Lambda}$ by automatically promoting rotations with many zero loadings thus resulting in many exact zeros in the factor loading matrix and facilitating identification. Differently from \cite{Knowles2011} and \cite{RD2008}, who do inference via a Gibbs sampler, \cite{Rockova2016} use an expectation-maximization (EM) algorithm, which brings computational advantages for high-dimensional data.

Recently, \cite{SFS2022} suggested an exchangeable shrinkage process (ESP) prior for finite number of factors $K$, which has relation to the IBP prior when $K \rightarrow \infty$. The prior in its general form is formulated as follows:
\begin{align} \label{ESP}
\lambda_{ih} \, | \, \tau_h \sim (1-\tau_h) \delta_0 + \tau_h P_{slab}(\lambda_{ih}), \qquad \tau_h \, | \, K \sim \mathcal{B}(a_K, b_K), \qquad h = 1, \ldots, K,
\end{align}
where $\delta_0$ is a Dirac delta, $P_{slab}$ is an arbitrary continuous slab distribution, and $K$ is the finite number of factors. The slab probabilities $\tau_h$s then decide the number of active factors $K_+ < K$. When in (\ref{ESP}) $b_K=1$ and $a_K=\alpha_{IB}/K$, for $K \rightarrow \infty$ this prior converges to the IBP prior (\cite{Teh2007}). The ESP prior has been used in the context of sparse Bayesian factor analysis in \cite{SFSDHHL2022} and in the context of a mixture of factor analysers model in \cite{MGSFS2023}.

\section{Generalised infinite factor models} \label{gen:inf:fac}

One of the recent developments in the area of infinite factor models is the generalised infinite factorisation model developed in \cite{Schiavon2022}, where authors were motivated by the existing methods' drawbacks such as lack of accommodation for grouped variables and other non-exchangeable structures. While the existing increasing shrinkage models focus on priors for $\bm{\Lambda}$ which are exchangeable within columns, they lack consideration for possible grouping of the rows of $\bm{\Lambda}$, which can occur in many applications, such as, for example, different genes in genomic data sets. Here we briefly outline the main idea of the proposed method without going into much detail.

The generalised model is defined in the following way:
\begin{align} \label{gen:mod}
y_{it} = s_i(z_{it}), \qquad \bm{z}_t = \bm{\Lambda} \bm{f}_t + \bm{\epsilon}_t, \qquad \epsilon_t \sim \eta_\epsilon,
\end{align}
where $\bm{\Lambda}$ is a $p \times K$ factor loading matrix, $\bm{f}_t$ is a $K$-dimensional factor with a diagonal covariance matrix $\bm{\Xi} = diag(\xi_{11}, \ldots, \xi_{KK})$, $\bm{\epsilon}_t$ is a $p$-dimensional error term independent of factors, $\eta_\epsilon$ is some arbitrary distribution, and the function $s_i$ is the function $s_i: \mathbb{R} \rightarrow \mathbb{R}$, for $i = 1, \ldots, p$. Here, differently from the factor model described in Section \ref{linBFA}, it is not necessarily assumed that $\bm{f}_t$ and $\bm{\epsilon}_t$ are normally distributed.

When, in fact, this is the case and $s_i$ is the identity function, the model (\ref{gen:mod}) takes the form of a Gaussian linear factor model described in Section \ref{linBFA}. When $s_i = F_i^{-1}(\Phi(z_{it}))$ with $\Phi(z_{it})$ denoting a Gaussian cumulative distribution function, the model (\ref{gen:mod}) becomes a Gaussian copula factor model as described in \cite{Murray2013}. Choosing an appropriate $s_i$ and modifying the assumptions regarding the distribution of the parameters in (\ref{gen:mod}) results in other types of factor models. The covariance matrix $\bm{\Omega}$ as in (\ref{eq:cov}) has a more general form in the case of the generalised infinite factorisation model $\bm{\Omega}  = \bm{\Lambda \Xi \Lambda}^T + \bm{\Sigma}$, where $\bm{\Sigma}$ is the covariance matrix of the error term.
The suggested prior on the elements of $\bm{\Lambda}$ allows infinitely many columns, so that the number of factors $K \rightarrow \infty$, and is formulated as follows:
\begin{align} \label{gen:lam}
\lambda_{ih} \, | \, \theta_{ih} \sim N(0,\theta_{ih}), \quad \theta_{ih} = \tau_0 \gamma_h \phi_{ih}, \quad \tau_0 \sim \eta_{\tau_0}, \quad \gamma_h \sim \eta_{\gamma_h}, \quad \phi_{ih} \sim \eta_{\phi_i},
\end{align}
where $\tau_0$, $\gamma_h$ and $\phi_{ih}$ are responsible for global, column-specific and local shrinkage, respectively, are independent a priori and the distributions $\eta_{\tau_0}$, $\eta_{\gamma_h}$ and $\eta_{\phi_i}$ are supported on $[0, \infty)$.

What is essentially different to previously described models, is that via $\phi_{ih}$ a non-exchangeable structure is imposed on the rows of $\bm{\Lambda}$ via some meta covariates $\bm{X}$, which inform the sparsity structure of $\bm{\Lambda}$. Denoting by $\bm{X}_{p \times q}$ a matrix of $q$ meta covariates, $\eta_{\phi_i}$ should be chosen so as to satisfy:
\begin{align*}
E(\phi_{ih} \, | \, \bm{\beta}_h) = g(\bm{x}_i^T \bm{\beta}_h), \quad \bm{\beta}_h=(\beta_{1h}, \ldots, \beta_{qh})^T, \quad \beta_{mh} \sim \eta_{\beta}, \quad m= 1, \ldots, q ,
\end{align*}
where $g$ is a smooth one-to-one differentiable link function, $\bm{x}_i = (x_{i1}, \ldots, x_{iq})$ denotes the $i$th row of $\bm{X}$, and $\bm{\beta}_h$ are coefficients controlling the impact of the meta covariates on the shrinkage of the elements of the $h$th column of $\bm{\Lambda}$. Taking the example from the ecology application studied in \cite{Schiavon2022}, different bird species (variables $i$) may belong to the same phylogenetic order (metacovariates $m$), have roughly the same size, follow similar diet etc.

In more details, the priors and hyperpriors on the factor loading are specified as follows:
\begin{eqnarray*}
&& \tau_0=1, \qquad \gamma_h = \nu_h \rho_h, \qquad \phi_{ih} \, | \, \bm{\beta}_h \sim Ber\{logit^{-1}(\bm{x}_i^T \bm{\beta}_h) c_p\}, \\
&& \nu_h^{-1} \sim \mathcal{G}(a_\nu, b_\nu), \quad a_\nu > 1, \quad \rho_h = Ber(1-\pi_h), \quad \bm{\beta}_h \sim N_q(0, \sigma^2_\beta \bm{I}_q),
\end{eqnarray*}
where the link function $g(x)$ takes the form of $logit^{-1}(x) = e^x/(1+e^x)$ and $c_p \in (0,1)$ is a possible offset. The distribution of the parameter $\pi_h = p(\gamma_h=0)$ follows a stick-breaking construction
\begin{align*}
\pi_h = \sum_{l=1}^h w_l, \qquad w_l = v_l \prod_{m=1}^{l-1} (1-v_m), \qquad v_m \sim \mathcal{B}(1, \alpha_{gen}),
\end{align*}
similar to \cite{LDD2020}. 

The model inference is performed via an adaptive Gibbs sampler, which resembles the one developed for the CUSP model. The frequency of adaptation is set in accordance with the Theorem 5 of \cite{Roberts2007}, and at the iteration, at which the adaptation occurs, the redundant columns of the loading matrix are discarded with all other corresponding parameters and the number of active factors is adapted accordingly. The redundant columns are identified as those for which $\rho_h=0$. If at some iteration there are no redundant columns, then an additional factor and all its corresponding parameters are generated from the priors.

The exact form of the Gibbs sampler steps depends on the prior assumptions for the elements of (\ref{gen:mod}). In case of the standard isotropic Gaussian and inverse gamma priors for factors and idyosyncratic variances, steps 2 and 3 of the sampler will be identical to the ones described in Section \ref{Gibbs:standard}. For the detailed description of the Gibbs sampler steps the reader is referred to the Supplementary Material of \cite{Schiavon2022}.

\section{Discussion and identification issues} \label{Conclusion}

Infinite factorisation models offer an enormous advantage of the automatic inference on the number of active factors by allowing it be derived from data. This is done by assigning a non-parametric prior to the elements of the factor loading matrix, which penalises the increasing number of columns. Some of such models at the same time account for the element-wise sparsity of factor loadings which can be justified in many real life applications, such as genetics, economics, biology, and many others.

One of the weak points of such models is that they often rely on rather subjective truncation parameters, with the lack of clear guidance towards the procedure of choosing such parameters. The MGP prior of \cite{BD2011} is the most prominent example of it, the simulation studies in \cite{Schiavon2020} and in Section \ref{simMGP} of this paper illustrate this point. This subjectivity was significantly reduced in the CUSP prior of \cite{LDD2020}. Generalisation of the CUSP prior by setting the hyperprior on the spike parameter as in \cite{Kowal2021} significantly improved the performance of the model on data sets of different nature and eliminated the need of data-dependent parameter tuning. In addition, the parameter-expanded version of the CUSP model suggested in \cite{Kowal2021} resulted in better uncertainty quantification. 

The class of generalised infinite factorisation models of \cite{Schiavon2022} generalises the idea of infinite factorisations with increasing shrinkage on factor loadings and incorporates it into a wide class of various types of factor models. In addition, it allows the grouping of the variables, which provides a useful feature for a wide rage of applications. The truncation of the redundant factors is done in a similar way to the CUSP model, however, the complexity of this rather general model makes unavoidable some subjective choices regarding hyperparameters and functional forms.

Another important issue concerns the identification of factor loadings. It is well known that the decomposition of the covariance matrix $\bm{\Omega}$ as in (\ref{eq:cov}) is not unique. First, the correct identification of the idiosyncratic covariance matrix should be ensured to guarantee that in the following two representations:
\begin{align*}
\bm{\Omega} = \bm{\Lambda} \bm{\Lambda}^T + \bm{\Sigma}, \qquad \quad \bm{\Omega} = \bm{\Theta} \bm{\Theta}^T + \bm{\Sigma}_0
\end{align*}
$\bm{\Sigma} = \bm{\Sigma}_0$ and, hence, the cross-covariance matrix $\bm{\Lambda} \bm{\Lambda}^T = \bm{\Theta} \bm{\Theta}^T$ is uniquely identified. This problem is known under the name of variance identification. The row deletion property of \cite{Anderson1956} presents a sufficient condition for variance identification and states that whenever an arbitrary row is deleted from $\bm{\Lambda}$, two disjoint matrices of rank $K$ should remain. This property imposes an upper bound on the number of factors $K \leq \frac{p-1}{2}$. So, for dense factor models, variance identification can fail if the number of factors is too high. For sparse factor models, additional restrictions on the number of non-zero elements in each column of $\bm{\Lambda}$ need to be applied (see, e.g. \cite{SFSDH2022}). Although in most cases $K \ll p$ and the upper bound will be respected, there is no formal guarantee of variance identification for infinite factor models even when the factor loading matrix is dense, and even less so in the case of sparse infinite factor models. 

The second problem deals with the correct identification of $\bm{\Lambda}$ from $\bm{\Lambda} \bm{\Lambda}^T$. It is referred to as the problem of rotational invariance and stems from the fact that for any semi-orthogonal matrix $\bm{P}: \bm{P} \bm{P}^T = \bm{I}$ and $\bm{\Theta} = \bm{\Lambda} \bm{P}$, $\bm{g}_t= \bm{P}^T \bm{f}_t$, the two models
    \begin{equation*}
           \bm{y}_{t} = 
          \bm{\Lambda} \bm{f}_t + \bm{\epsilon}_{t}  \qquad \text{and} \qquad \bm{y}_{t} = \bm{\Theta} \bm{g}_t + \bm{\epsilon}_{t}
    \end{equation*}
are observationally indistinguishable. This problem is often addressed in the literature by imposing restrictions on the elements of $\bm{\Lambda}$, such as, for example, setting the upper diagonal elements equal to zero and requiring the diagonal elements to be positive so that $\bm{\Lambda}$ represents a positive lower triangular matrix. This approach has first been implemented by \cite{GZh1996} and followed by many others (see, for example, \cite{Lopes2004} and \cite{Carvalho2008}). This constraint introduces order dependence upon variables, which results in posterior distributions whose shapes depend on the ordering of the variables in the data set and thus is not applicable for infinite factor models. However, these models can still be employed for the tasks of covariance matrix estimation, variable selection and prediction, which do not require identification.

However, while variance identification is rarely addressed in the literature and not at all in the context of infinite factor models, in recent years some ex-post identification methods aimed at tackling rotational invariance have been proposed, which are applicable for infinite factor models. These methods usually involve some kind of orthogonalisation procedure applied at a post-processing step, such as, for example, orthogonal Procrustean algorithm (\cite{Ausmann2016}) or Varimax procedure (\cite{PFD2021}).

There have also been some attempts to embed identification consideration into the estimation procedure. Thus, \cite{Rockova2016} offer a solution to the indeterminacy due to rotational invariance via the SSL prior, which automatically promotes the rotations with many zero loadings and thus reduces posterior multimodality. Their EM algorithm provides sparse posterior modal estimates with exact zeroes in the factor loading matrix. \cite{Schiavon2022} propose an identification scheme, which is somewhat similar in the idea. They search for an approximation of the maximum a posteriori estimators of $\bm{\Lambda}$, $\bm{\beta}=(\beta_1, \beta_2, \ldots)$ and $\bm{\Sigma}$ by integrating out the scale parameters and latent factors from the posterior density function and taking the parameters of interest from the draw which produced the highest marginal posterior density function $f(\bm{\Lambda}, \beta,\bm{\Sigma} \, | \, \bm{y})$.

\clearpage
\nocite{}
\setlength\bibitemsep{1.2\itemsep}
\printbibliography

\end{document}